\documentclass{AGUJournal}
\draftfalse

\journalname{Geophysical Research Letters}

\usepackage[utf8]{inputenc}
\usepackage[amsmath,upgreek]{}

\begin{document}
	
	
	\title{Seasonal Water ``Pump'' in the Atmosphere of Mars: Vertical Transport to the Thermosphere}
	
	
	\authors{Dmitry S. Shaposhnikov\affil{1,2}, Alexander S. Medvedev\affil{3}, Alexander V. Rodin\affil{1,2}, and Paul Hartogh\affil{3}}
	
	\affiliation{1}{Moscow Institute of Physics and Technology, Moscow, Russia}
	\affiliation{2}{Space Research Institute, Moscow, Russia}
	\affiliation{3}{Max Planck Institute for Solar System Research, G\"ottingen, Germany}
	
	\correspondingauthor{D. S. Shaposhnikov}{shaposhnikov@phystech.edu}
	
	
	\begin{keypoints}
		\item Global circulation modeling reveals the mechanism of water exchange between the lower and upper atmosphere
		\item Atmospheric dust controls the circulation strength and, hence, the amount of high-altitude water
		\item Solar tide modulates the upwelling of water vapor by almost completely shutting it down during certain local times
	\end{keypoints}
	
	
	\begin{abstract}
		We present results of simulations with the Max Planck Institute general circulation model (MPI--MGCM) implementing a hydrological cycle scheme. The simulations reveal a seasonal water ``pump'' mechanism responsible for the upward transport of water vapor. This mechanism occurs in high latitudes above 60$^\circ$ of the southern hemisphere at perihelion, when the upward branch of the meridional circulation is particularly strong. A combination of the mean vertical flux with variations induced by solar tides facilitates penetration of water across the ``bottleneck'' at approximately 60~km. The meridional circulation then transports water across the globe to the northern hemisphere. Since the intensity of the meridional cell is tightly controlled by airborne dust, the water abundance in the thermosphere strongly increases during dust storms.
	\end{abstract}
	
	
	\section{Introduction}
	
	Water is a minor component of the Martian atmosphere, which is largely confined within a few lower scale heights. Nevertheless, it is the main source of hydrogen in the upper atmosphere \citep{hunten1970production, parkinson1972spectroscopy, krasnopolsky2002mars}. Escape of hydrogen atoms into space near the exobase varies by an order of magnitude seasonally, maximizing around southern summer solstice (solar longitude $L_s\approx270^\circ$), according to MAVEN \citep{halekas2017seasonal} and HST observations \citep{bhattacharyya2017seasonal} during dust storms \citep[e.g.,][]{bhattacharyya2015strong, chaffin2014unexpected, clarke2014rapid, clarke2017variability}. Observed water in the lower atmosphere also experiences strong seasonal changes and depends on airborne dust load \citep[e.g.,][]{smith2009compact, maltagliati2011annual, trokhimovskiy2015mars, pottier2017unraveling}. This implies a link between water in the troposphere and thermosphere and a corresponding mechanism of transport between the layers.
	
	The Martian middle atmosphere is too cold to sustain water vapor, especially around the mesopause, while ice particles are sufficiently heavy and prone to sedimentation. This water behavior is similar to that in the terrestrial middle atmosphere \citep{Seele1999}. However, there are multiple observations showing a presence of water vapor in the middle atmosphere at certain locations and times \citep[e.g.,][]{maltagliati2013annual, fedorova2018water}. \citet{heavens2018hydrogen} and \citet{fedorova2018water} provided evidence of strong seasonal variations of the globally averaged water abundance and its vertical extension up to 70-80~km at perihelion during the Martian Year 28 (MY28) global dust storm. Hypotheses concerning the mechanism of vertical transport of water include mesoscale deep convection \citep{heavens2018hydrogen}, turbulent mixing in the lower atmosphere and/or an unspecified dynamics in the upper atmosphere \citep{clarke2018nature}. General circulation modeling underestimates the hygropause altitude at southern summer solstice to date \citep{chaufray2015variability, pottier2017unraveling}.   
	
	Our study addresses this gap in knowledge of processes that couple water in the lower and upper atmosphere. We present results of simulations with our recently developed hydrological scheme \citep{shaposhnikov2018modeling} implemented in the Max Planck Institute Martian general circulation model (MPI--MGCM). This is the first modeling study that considers in detail the transport of water from the surface to the thermosphere and explores its dependence on dust storms. In section~\ref{sec:MGCM}, we outline the modeling tools and setup of numerical experiments. The annual cycle of vertical water transport is discussed in Section~\ref{sec:flux}. In sections~\ref{sec:perihelion} and \ref{sec:local-time}, we zoom in on the perihelion season and explore the zonal mean transport of water and local time variations, correspondingly. The results of simulations are compared with observations from Mars Climate Sounder onboard Mars Reconnaissance Orbiter (MCS--MRO) in section~\ref{sec:mcs}.
	
	\section{Model description and Design of Simulations}
	\label{sec:MGCM}
	
	The MPI--MGCM has been described in detail in the papers of \citet{hartogh2005description, hartogh2007middle, medvedev2007winter}. It employs a spectral dynamical solver for the primitive equations of hydrodynamics on a sphere. In the vertical, the grid is discretized by 67 hybrid $\eta$-levels, terrain-following near the surface and pressure based near the top at $3.6\times 10^{-6}$ Pa ($\sim$160~km). The model includes a set of physical parameterizations suitable for the Martian atmosphere from the ground to the thermosphere. T21 horizontal resolution (corresponding to $\approx 5.6^\circ$) was used in the simulations. Subgrid-scale gravity waves were parameterized as described in the work of \citet{medvedev2011influence}. 
	
	The hydrological part of the model has been described in detail in the papers of \citet{shaposhnikov2016water, shaposhnikov2018modeling}. It includes a semi-Lagrangian transport of water vapor and ice, and accounts for the microphysics of vapor-ice conversions. Ice clouds are formed whenever water vapor condenses on cloud condensation nuclei (CCN). The heterogeneous nucleation rate and ice particle growth rate are evaluated according to \citet{jacobson2005fundamentals}. The sizes of CCN are represented by four bins. A two-moment scheme is applied to every bin for separately keeping track of the ice mass and number of particles \citep{rodin2002moment}. The size of ice particles determines their microphysical properties and the sedimentation rate. The CCN number density in each bin is calculated from the bimodal log-normal dust distribution \citep{fedorova2014evidence}, as described in the paper of \citet[][section 2.2]{shaposhnikov2018modeling}.
	
	Unlike with our previous simulations of the water cycle \citep{shaposhnikov2018modeling}, those presented here have been performed in the domain extending into the thermosphere, where water is no longer chemically conservative and accurate photochemical modeling may have to be included depending on the motive \citep[e.g.,][]{hartogh2010alomar}. For purposes of this work, we retained only a parameterization of H$_2$O losses due to photodissociation. The water photodissociation rates have been calculated according to \citet[][formulae (1)--(2)]{anbar1993photodissociation}.
	
	In this study we employ two predetermined dust scenarios. The ``basic'' one represents an observationally-based seasonally and latitudinally evolving (i.e., zonally averaged) aerosol optical depth $\tau$ in the thermal IR based on MGS-TES and MEX-PFS measurements with the global dust storms removed \citep{medvedev2011influence}. The second one is based on the measurements for the Martian Year 28 (MY28), which included a major dust storm during the perihelion season \citep[][Figure~3]{medvedev2013general}. In both scenarios, vertical profiles of dust were prescribed after \citet{Conrath75} with modifications described in the paper of \citet[][Formulae 1 and 2]{medvedev2013general}.
	
	The model has been initialized with the distribution of water vapor and ice obtained in our earlier simulations \citep{shaposhnikov2018modeling}. The latter runs have been performed for several Martian years until the model achieved a quasi-stable state. Since the current version of the MGCM extends higher into the thermosphere, the additional vertical levels have been initialized with the values at $\sim$100~km. The initial conditions for the dynamical fields are taken from the simulations of \citet{medvedev2016comparison}.
	
	The total amount of water in the atmosphere depends on the model time step due to the instability of commonly used nucleation and particle growth schemes \citep[][see the discussion around Figure 11]{navarro2014global, shaposhnikov2018modeling}. Therefore, we applied a 10 s time step for microphysics and other model processes in order to suppress instabilities and increase accuracy.  
	
	\section{Vertical Transport of Water Vapor}
	\label{sec:flux}
	
	Vertical transport of water vapor and ice is best characterized by the corresponding fluxes. Figure~\ref{fig:flux-slices} presents latitude-seasonal distributions of the vertical water vapor flux at several altitudes simulated using the ``basic'' dust scenario. It  clearly shows that at all altitudes above 30~km, the flux maximizes around perihelion between $L_s=200^\circ$ and 300$^\circ$ (Figure~\ref{fig:flux-slices}b-f) and is negligibly small throughout the rest of the year. The flux distributions are approximately symmetric with respect to the solar longitude $L_s=260^\circ$, when the global mean temperature reaches its annual maximum. Between $L_s=220^\circ$ and 300$^\circ$, the transport of water vapor up to $\sim$90~km follows the meridional circulation cell with air rising in the summer hemisphere and sinking in the winter one. In the thermosphere above 120~km, the distributions indicate additional circulation cells in low-to-middle latitudes. However, the pole-to-pole transport persists, and the magnitudes of upward and downward fluxes over the, correspondingly, southern (summer) and winter (northern) poles significantly increase. 
	
	The other thing that stands out in Figure~\ref{fig:flux-slices}c is the $\sim$2~ppmv~m~s$^{-1}$ minimum of the upward water flux at perihelion located at around 60~km. This is the only region (between 20$^\circ$S and 70$^\circ$S) where at certain times of the year water can penetrate from the lower atmosphere into the upper layers. Once water is through this ``bottleneck'', it is transported further upward and across latitudes northward. Note that not all the water stays in the upper atmosphere with the prospect of being photolyzed and ultimately escaping to space. The global circulation also returns a portion of water to the lower atmosphere in the northern polar region, as is shown with bluish shades in Figure~\ref{fig:flux-slices}c.
	
	\section{Water Transport at Perihelion}
	\label{sec:perihelion}
	
	We next zoom in on the perihelion season and consider the water transport in more detail. Figure~\ref{fig:perihelion}a presents the water vapor amount in ppmv averaged diurnally and between $L_s$=$250^\circ$ and $270^\circ$. Streamlines show the residual meridional circulation, while their thickness and color indicate the magnitude and vertical direction of the water vapor flux, correspondingly. In agreement with previous observations and simulations, it is seen that the dominant part of water vapor concentrates in the southern (summer) hemisphere below $\sim$45~km. Water increasingly sublimates near the surface at middle to high latitudes and is transported up- and northward by the meridional cell. This results in the water vapor maximum of up to a few hundred ppmv at around 30~km that extends in latitude to $\sim$45$^\circ$N. Water ice clouds (shown with white contours) form immediately above and are transported by the meridional circulation in the same manner as vapor. Color shades in Figure~\ref{fig:perihelion}a demonstrate an elevated amount of water vapor ($\sim$90--140~ppmv) in the high-latitude ``bottleneck'' between $\sim$60 and 90~km. Higher up at around the mesopause and in the lower thermosphere (see Figure~\ref{fig:perihelion}c), the water vapor is effectively transported across the globe by the meridional circulation, and its magnitude increases up to $\sim$160~ppmv (in the average sense mentioned above). 
	
	Thick contour lines in Figure~\ref{fig:perihelion}a demonstrate a strong downward flux of vapor in the north polar region at all altitudes in the middle and upper atmosphere. As a consequence, the water vapor mixing ratio increases there as well. Note that the same water mass produces larger volume mixing ratio in the upper atmosphere due to exponentially decaying pressure and density. This downwelling over the winter pole is the major mechanism of returning water back to the lower atmosphere. Due to colder temperature and higher pressure, this water condenses and contributes to the ice polar hood at around 30~km (depicted by the white contour lines). Unlike molecular diffusion, which prevents accumulation of water in the upper atmosphere at all latitudes and times, this mechanism is distinctively seasonal and localized. 
	
	\section{Water During Dust Storm}
	\label{sec:dust-comparison}
	
	After illustrating the mechanism of lifting water vapor up through the ``bottleneck'', we turn our attention to the causal relationship between the high-altitude water and atmospheric dust. There are several channels, through which the former can affect the latter. Absorption of solar radiation by airborne aerosol increases temperature and, thus, hinders condensation of water. On the other hand, the increasing number of particle nuclei facilitates formation of ice clouds. Finally, dust storms strongly amplify the meridional circulation \citep[e.g.,][]{Medvedev2011dust}. 
	
	We repeated the simulations with the dust scenario based on column opacity measurements during MY28 when a major planet-encircling dust storm occurred around the season of interest. For simplicity, we use the direct model output of water vapor rather than applying the indirect method of \citet{heavens2018hydrogen} to the model temperature and water ice fields. Figure~\ref{fig:perihelion}d shows that temperature increased by $\sim$20~K over the south pole and by more than 30~K over the north pole at 45~km (see color contours for temperature differences with the ``basic'' dust scenario). The meridional transport intensified during the dust storm. In particular, the warming over the winter pole is caused adiabatically by the downward branch of the circulation cell \citep[e.g.,][]{hartogh2007middle, Kuroda_etal2009}. The changes in temperature and transport remarkably affected atmospheric water. Thus, the total amount of water vapor in the atmosphere noticeably increased (see Figure~\ref{fig:perihelion}b) and its upper boundary (hygropause) extended above $\sim$60~km, $\sim$10~km higher than in the ``basic'' dust scenario. Correspondingly, the simulated ice clouds became denser, and their top has elevated by $\sim$10~km in the southern hemisphere. The enhancement of the circulation is the global phenomenon that covers altitudes up to the thermosphere. Consequently, the vertical transport of water vapor increases in the southern hemisphere as well, which is depicted by thicker contour lines in Figure~\ref{fig:perihelion}b. This produced a maximum of water vapor of up to 220~ppmv between 120 and 150~km at 30$^\circ$S--90$^\circ$S. Somewhat warmer temperatures in low to middle latitudes in the mesosphere and thermosphere during the dust storm (see color contours in Figure~\ref{fig:perihelion}d) along with the enhanced transport produced increased vapor abundances at all heights above $\sim$70~km. It is seen that there are fewer ``dry'' regions in the middle atmosphere compared to those for the ``basic'' dust simulation. Finally, the simulation reproduces the change in water vapor abundance in the south observed by \citet{fedorova2018water} but seems to underestimate water vapor at 60--75~km in the tropics and northern extratropics (as observed by both \citet{fedorova2018water} and \citet{heavens2018hydrogen}).
	
	Another effect of the dust storm captured by the model is the increase of CCN in the atmosphere. A larger number of nuclei aids water vapor condensation, the formed ice particles have smaller radii and, thus, slower sedimentation speed. Therefore, water ice clouds form higher, which too contributes to increased water abundances in the upper atmosphere. 
	
	We illustrated that an increase of the airborne dust ``widens'' up the ``bottleneck'' for water penetration into the upper atmosphere in the high-latitude southern hemisphere both dynamically and microphysically. The above consideration was based on the mean (diurnally/zonally averaged) fields. In the next section, we turn our attention to local time variations.
	
	\section{Local Time Variations}
	\label{sec:local-time}
	
	Figure~\ref{fig:local-time} shows deviations of the water vapor abundance (shaded) and vertical velocity (contours) from the corresponding zonal mean quantities as a function of local time in a particular grid point close to the south pole (75$^\circ$S, 0$^\circ$E). This composite plot is based on multi-day averaging over the period between $L_s=250^\circ$ and $270^\circ$. It is seen from Figure~\ref{fig:local-time}a for the ``basic'' dust scenario that the vertical velocity exhibits a mixture of the diurnal and semidiurnal tides with phase advancing with height. The downward phase tilt is the manifestation of the tide generated below and propagating upward. The angle of the tilt progressively changes with altitude above $\sim$70~km to almost vertical, indicating the increasing role of the in situ-excited tides in the upper atmosphere. Note that our results are shown for high latitudes of the summer hemisphere (above 75$^\circ$S), whereas modeling \citep[e.g.,][]{ForbesMiyahara2006} and MCS--MRO observations \citep{Kleinb-semidiurnal2013} provide evidence that the amplitude of the semidiurnal temperature variations maximizes in middle- to high latitudes of the winter hemisphere. There is no contradiction in that, because tide is a global phenomenon, in which amplitudes of fluctuations of different field variables can peak at different latitudes \citep[e.g.,][]{Yigit_etal2017tide}. 
	
	Unlike the vertical velocity, water vapor varies mainly with the diurnal periodicity with the maximum magnitude of $\sim$120~ppmv at 30~km. Interactions with the semidiurnally varying vertical velocity form a characteristic steep reversal of water anomalies at 40~km by pushing water up and down twice per day \citep[cf. to Earth,][]{hallgren2012}. The magnitude of the diurnal variations of water vapor at the ``bottleneck'' altitude around 60~km is 30--50~ppmv. Given that the mean vapor abundance in this region is around 70~ppmv, the total amount varies considerably with more water during the first half of a day. Higher in the middle and upper atmosphere, temporal variations of vapor are smaller, but correlate more with the upward fluxes. 
	The major dust storm contributes to formation of giant diurnal water vapor variations in the lower atmosphere of greater than 500~ppmv at 30--40~km (Figure~\ref{fig:local-time}b). They occur due to enhanced sublimation from the reservoir on/under the surface. The variations extend higher into the middle atmosphere with the magnitude of $\sim$100--150~ppmv at the ``bottleneck'' around 60~km. A comparison with the zonal mean values in Figure~\ref{fig:perihelion}b shows that some ``leakage'' of vapor into the middle and upper atmosphere takes place also during the second half of day in addition to strong tidally-modulated pulses during the first half. Above the mesopause, the semidiurnal tide weakens during the dust storm, and its correlation with the water vapor amount becomes less certain. 
	
	\section{Comparison With MCS Observations}
	\label{sec:mcs}
	
	In order to validate our simulations, we compare them with the data inferred from the measurements by the MCS--MRO during MY28 \citep{heavens2018hydrogen}. The authors of the latter study used an indirect method to estimate the water vapor and ice abundances from the observations of temperature and water ice clouds. 
	MCS performs 13 polar orbits per Martian sol. Away from the poles, the groundtrack of MRO corresponds to approximately 15:00 hours local solar time on the ascending side of the orbit (Figure~\ref{fig:mcs}a) and to $\sim$3:00 local time on the opposite side (Figure~\ref{fig:mcs}c). Since MCS orbits vary, we used the model output averaged over the intervals 14:00--16:00 and 2:00--4:00 hours local time for comparison, correspondingly. 
	Both observations and modeling in Figure~\ref{fig:mcs} show gradual, but rapid increase of the total water abundance and its rise in altitude towards the perihelion season. There were no successful retrievals available during the dust storm itself between approximately $L_s = 260^\circ$ and $305^\circ$ and above 80~km. However, at the highest available levels between 70 and 80~km, both the model and observations agree well in showing $\sim$70--80~ppmv of water. Good agreement also exists with the night-time measurements at 40--50~km immediately before the onset of the dust storm ($L_s=220^\circ$ to 260$^\circ$). Greater water abundances during the night time at these altitudes are due to the tidal phase (higher temperature), as can be seen from Figure~\ref{fig:local-time}b. Note that the MCS measurements demonstrate that the maxima of water are vertically localized around 40--50 km between $L_s=200^\circ$ and 250$^\circ$, whereas in the model the water mixing ratio increases down to the surface. This difference may be due to the adopted dust scenario that does not capture detached dust layers. The model also demonstrates a rapid fall of water abundances after $L_s = 330^\circ$, which is not supported by the observations.
	
	Nevertheless, the total observed and simulated amount of water during the dust storm as well as the shape of the seasonal distribution agree well, at least in the latitudinally averaged sense presented here. They clearly illustrate that conditions for upward water penetration across the ``bottleneck'' at $\sim$60~km exist only during a limited time of the year around perihelion, and dust storms strongly enhance this penetration.
	
	\section{Discussion and Conclusions}
	\label{sec:conclusions}
	
	Several mechanisms have been proposed to explain the observed presence of water in the middle atmosphere of Mars above 60~km. \citet{maltagliati2011evidence} suggested supersaturation of water vapor due to purely microphysical reasons (lack of condensation nuclei). \citet{clarke2018nature} considered the dynamics and hypothesized that either turbulent mixing in the lower atmosphere raises water vapor upward, or the strengthened by solar UV circulation in the upper atmosphere facilitates this transport. \citet{heavens2018hydrogen} attributed the appearance of water vapor and ice at upper levels to deep convection enhanced dust storms. Our simulations with the general circulation model revealed the full picture of water transport from the ground up to the thermosphere. The main findings are the following.  	
	\begin{itemize}
		\item Water is lifted up in high latitudes of the summer hemisphere by the upward branch of the pole-to-pole meridional circulation cell. It is then transported by the latter across latitudes in the mesosphere and thermosphere.
		\item Water can penetrate upper levels only during the perihelion season, when the meridional circulation cell is sufficiently strong.
		\item The influx of water to the middle and upper atmosphere increases, whenever the meridional cell intensifies, for instance, during dust storms. In addition, dust storm-induced heating increases the amount of water vapor in the lower atmosphere.
		\item Upward transport of water is significantly modulated by the solar tide. The latter acts as a ``pump'' by increasing the transport during certain local times and almost completely shutting it down during the others.
	\end{itemize}
	
	The described transport of water to the Martian upper atmosphere has some similarities and differences with that on Earth. In the terrestrial stratosphere and mesosphere, there is also a strong water upwelling in the summer hemisphere that even compensates for photochemical destruction \citep{hartogh2010alomar}. Due to the circular orbit and unlike on Mars, it occurs during both solstices. However, water in the terrestrial atmosphere is rapidly destroyed by photolysis in the sun-lit summer hemisphere below 70~km, whereas on Mars its significant portion can be transported across the globe. 
	
	Photochemical calculations \citep{chaffin2017elevated, krasnopolsky2019escape} suggest that water abundances of $\sim$80~ppmv at 60--80~km can explain the observed magnitudes of hydrogen escape at the exobase. Our simulations show that, even at dustless seasons, the circulation can deliver these amounts of water over the southern high latitudes, at least during certain local times. Moreover, comparable abundances of vapor are distributed by the circulation over all latitudes above $\sim$120~km. During major dust storms (similar to that of MY28), the corresponding water abundances increase by a factor 2 and more. Overall, our simulations at least partly reconcile the existing observations and estimates, reveal the impact of planetary-scale circulations on the behavior of water in the middle and upper atmosphere, and provide testable predictions for evaluating alternative hypotheses against future observations.
	
	\acknowledgments
	
	The data supporting the MPI-MGCM simulations can be found at \underline{https://\-mars.mipt.ru}, \underline{https://\-zenodo.org/record/1553514} \citep{shaposhnikov_d_s_2018_1553514} or obtained from D. Shaposhnikov (shaposhnikov@phystech.edu).
	
	The authors thank Anna A. Fedorova, Takeshi Kuroda and Nicholas Heavens for assistance with the observational data and helpful discussions. This work has been performed at the Laboratory of Applied Infrared Spectroscopy of Moscow Institute of Physics and Technology and at Max Planck Institute for Solar System Research. The work was partially supported by the Russian Science Foundation grant 16-12-10559 and German Science Foundation (DFG) grant HA3261/8-1.
	
	
	

	
	\listofchanges
	
	\begin{figure}[ht]
		\setcounter{figure}{0}
		\centerline{\includegraphics[width=40pc]{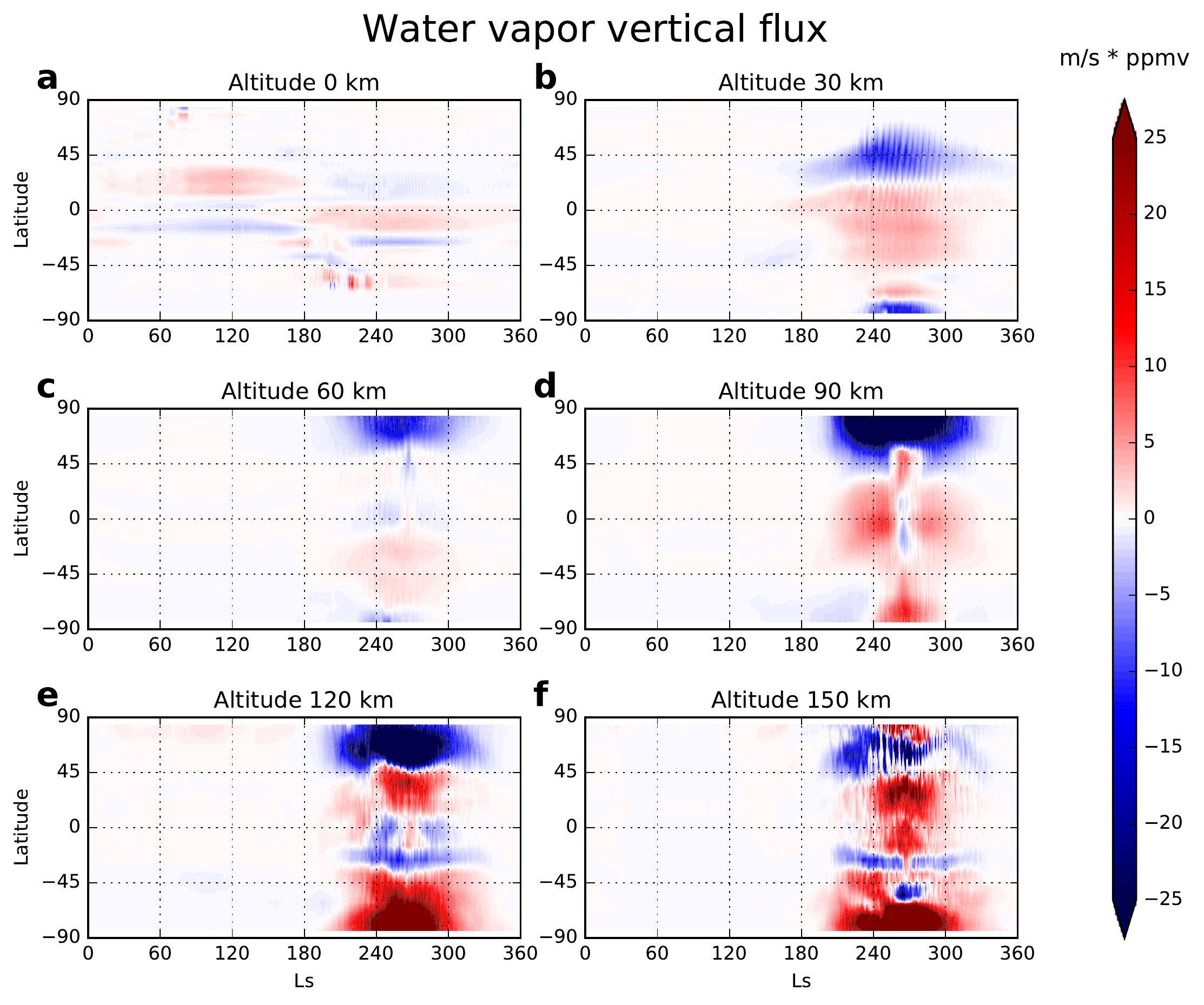}}
		\caption{Latitude-seasonal variations of the zonally averaged vertical water vapor flux simulated using the ``basic'' dust scenario (see section~\ref{sec:MGCM}) at different altitudes: 0, 30, 60, 90, 120 and 150~km (panels a to f, correspondingly). Positive values (upward fluxes) are plotted in red, negative (downward) fluxes are shown in blue.}
		\label{fig:flux-slices}
	\end{figure}
	
	\begin{figure}[ht]
		\setcounter{figure}{1}
		\centerline{\includegraphics[width=40pc]{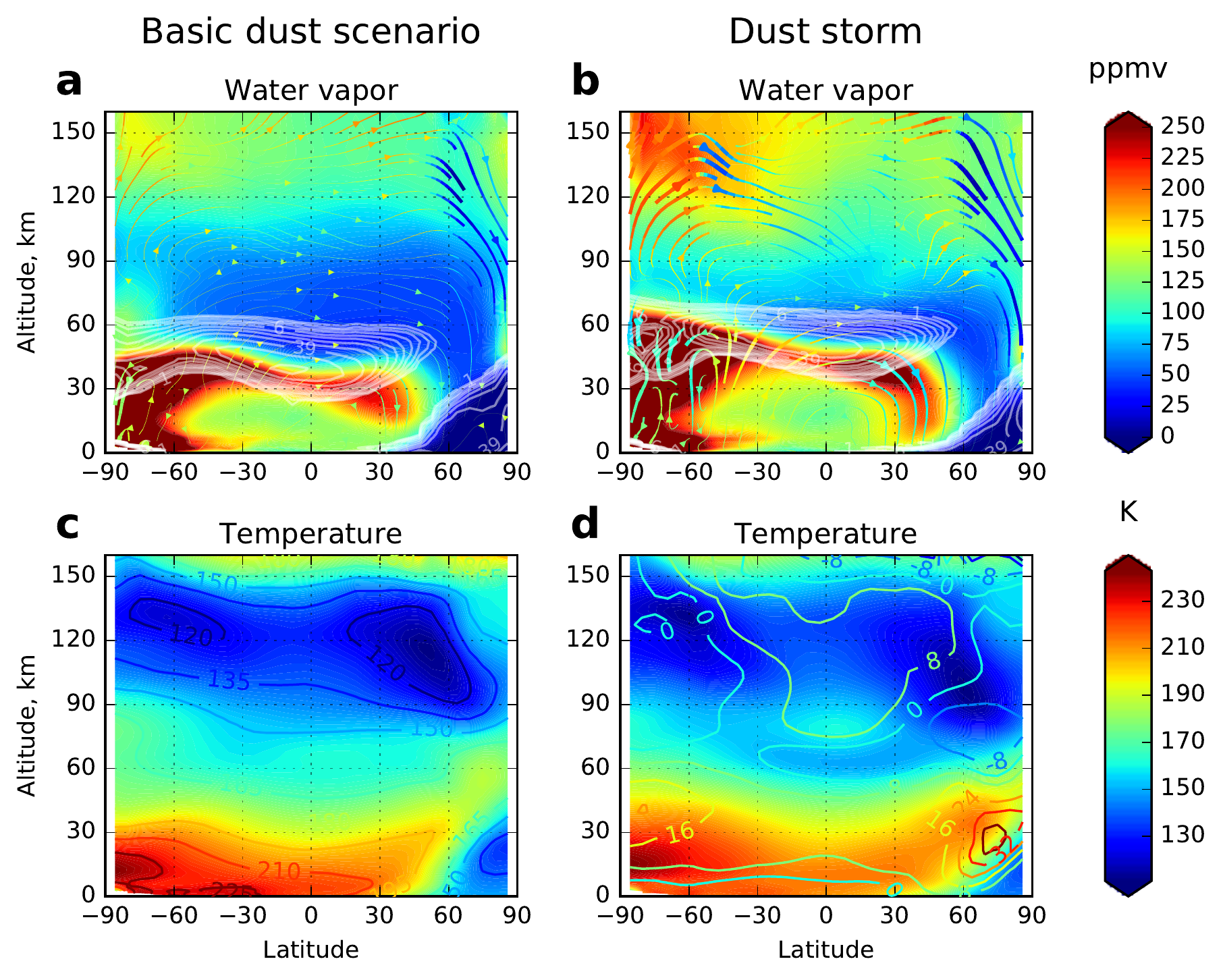}}
		\caption{Latitude-altitude cross-sections of the quantities simulated for the ``basic'' dust scenario (left column) and the MY28 dust storm (right column):
			(a) Water vapor (shaded), water ice (white contours) and the meridional flux of water vapor (the lines with arrows, the color and thickness of which indicate the vertical direction and magnitude, correspondingly); 
			(b) is the same as in panel (a), but for the dust storm of MY28;
			(c) temperature (shaded) for the ``basic'' dust scenario; 
			(d) is the same as in (c), but for the MY28 dust storm scenario, except for the contour lines that show the temperature difference between (d) and (c). All fields are averaged zonally and over the period between $L_s=250^\circ$ and $270^\circ$. }
		\label{fig:perihelion}
	\end{figure}
	
	\begin{figure}[ht]
		\setcounter{figure}{2}
		\centerline{\includegraphics[width=40pc]{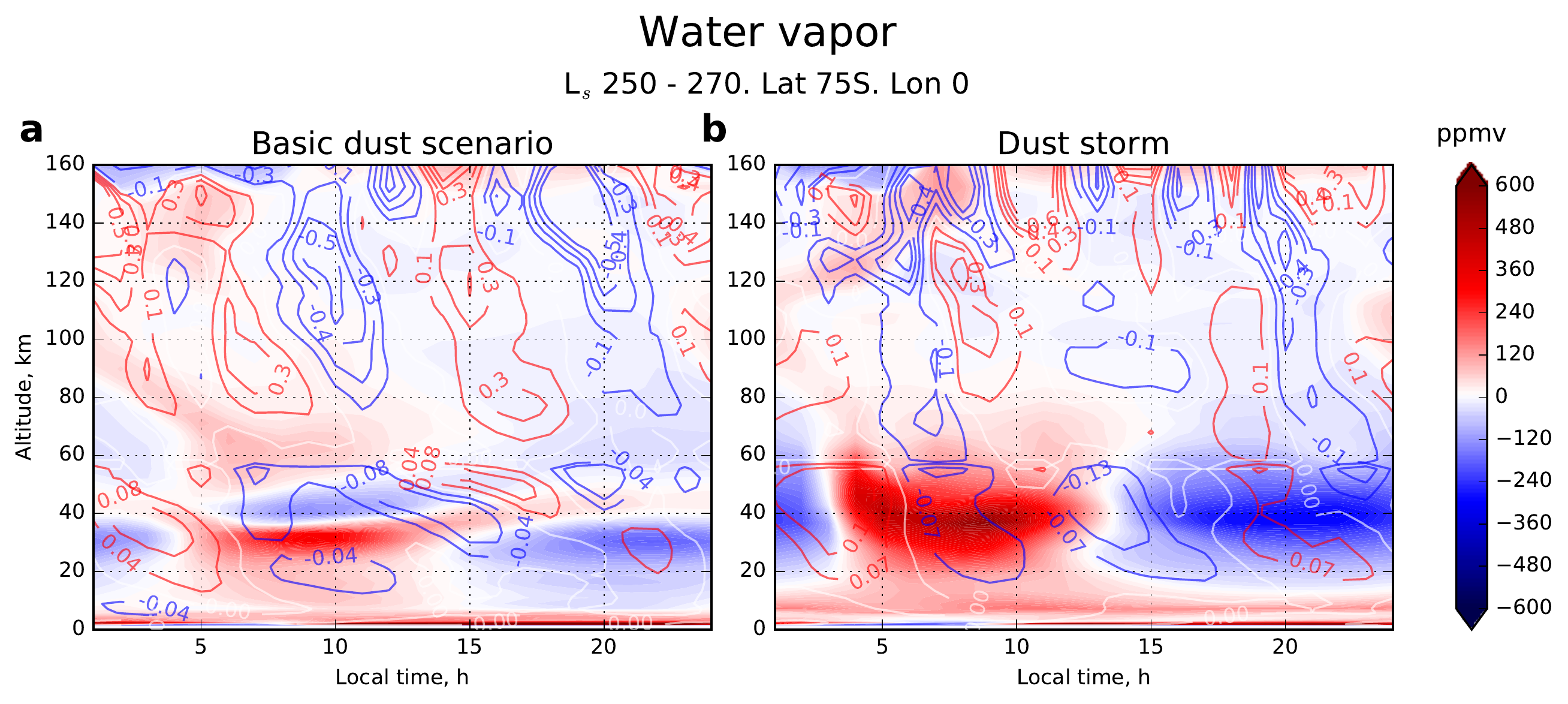}}
		\caption{Altitude-local time distributions of deviations from the zonal mean for water vapor (in ppmv, color shades) and vertical velocity (in m~s$^{-1}$, contours). Shown are the composite over the period between $L_s=250^\circ$ and $270^\circ$ as functions of local time close to a grid point located near 75$^\circ$S and 0$^\circ$ longitude. Positive values of the vertical velocity correspond to upward motions. Panel (a) and (b) are for the simulations with the ``basic'' and ``MY28 dust storm'' dust scenarios, correspondingly.  }
		\label{fig:local-time}
	\end{figure}
	
	\begin{figure}[ht]
		\setcounter{figure}{3}
		\centerline{\includegraphics[width=40pc]{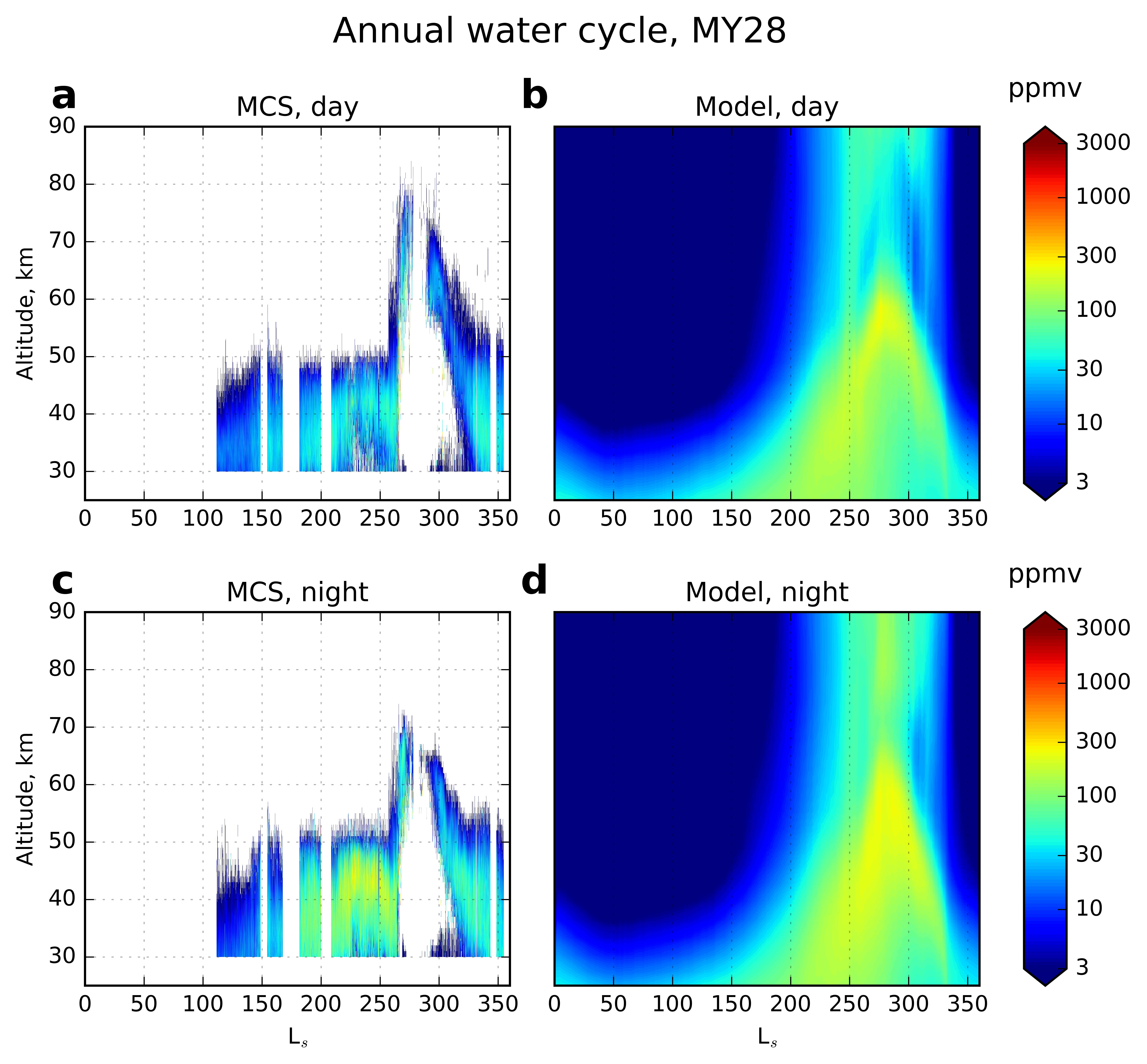}}
		\caption{Vertical distribution of the total water (vapor+ice) content derived from the Mars Climate Sounder (MCS) measurements (left column) \citep{heavens2018hydrogen} and simulated with the MPI--MGCM (right column) for the MY28: for the day side ($\sim$15:00 local time, upper row) and night side (03:00 local time, lower row). In all panels, the values were averaged over longitudes and latitudes. In the simulations, the averaging over 14:00--16:00 and 02:00--04:00 local times was performed.}
		\label{fig:mcs}
	\end{figure}
	
\end{document}